\documentstyle[12pt]{article}

\catcode`\@=11 \@addtoreset{equation}{section}\catcode`\@=12
\newcommand{\be}{\begin{equation}}
\newcommand{\ee}{\end{equation}}
\newcommand{\ba}{\begin{eqnarray}}
\newcommand{\ea}{\end{eqnarray}}

\begin{document}
\thispagestyle{empty}

\begin{center}
               RUSSIAN GRAVITATIONAL ASSOCIATION\\
               CENTER FOR SURFACE AND VACUUM RESEARCH\\
               DEPARTMENT OF FUNDAMENTAL INTERACTIONS AND METROLOGY\\
\end{center}
\vskip 4ex
\begin{flushright}                              RGA-CSVR-014/94\\
                                                gr-qc/9411xxx
\end{flushright}
\vskip 15mm

\begin{center}

{\large\bf  Billiard Representation for Multidimensional Quantum Cosmology
near the Singularity}

\vskip 5mm
{\bf
V. D. Ivashchuk and V. N. Melnikov }\\
\vskip 5mm
     {\em Center for Surface and Vacuum Research,\\
     8 Kravchenko str., Moscow, 117331, Russia}\\
     e-mail: mel@cvsi.uucp.free.net \\
\end{center}
\vskip 10mm

ABSTRACT

The degenerate Lagrangian system  describing
a lot of cosmological models is
considered. When certain restrictions on the parameters of the model are
imposed, the dynamics of the model near the "singularity" is reduced
to a billiard on the  Lobachevsky space. The Wheeler-DeWitt equation
in the asymptotical regime is solved and a third-quantized model
is suggested.

\vskip 10mm

PACS numbers: 04.20, 04.40.  \\

\vskip 30mm

\centerline{Moscow 1994}
\pagebreak

\setcounter{page}{1}

\pagebreak

\section{Introduction}
\setcounter{equation}{0}

In the present paper we deal with a stochastic behavior in
multidimensional cosmological models near the singularity (see [1-8]
and references therein). A large
variety of these models may be described by the following Lagrangian [9]
\begin{equation}
L = {L}(z^{a}, \dot{z}^{a}, {\cal N}) = \frac{1}{2} {\cal N}^{-1}
\eta_{ab} \dot{z}^{a} \dot{z}^{b} -  {\cal N} {V}(z),
\end{equation}
where ${\cal N} >0 $ is the Lagrange multiplier (modified lapse function),
$(\eta_{ab})=  diag(-1,+1, \ldots ,+1)$  is matrix of minisuperspace
metric, $a,b = 0, \ldots , n-1$, and
\begin{equation} 
{V}(z) = A_0 + \sum_{\alpha=1}^{m} A_{\alpha} \exp(u^{\alpha}_a z^a)
\end{equation} is the potential.
$A_0 >0$  corresponds to Zeldovich matter and $A_{\alpha} \neq 0$.
We impose the following restrictions on the vectors $u^{\alpha}
= (u^{\alpha}_0, \vec{u}^{\alpha})$
in the potential (1.2)
\begin{eqnarray} 
&& 1) A_{\alpha} > 0, \ if \ (u^{\alpha})^2 = -(u^{\alpha}_0)^2 +
(\vec{u}^{\alpha})^2 > 0;         \\
&& 2) u^{\alpha}_0 > 0  \ for \ all \  \alpha  = 1, \ldots, m.
\end{eqnarray}

Here we consider the classical and quantum behavior of the dynamical system
(1.1) for $n \geq 3$ in the limit
\begin{equation} 
z^2  = -(z^0)^2 + (\vec{z})^2   \rightarrow  -\infty, \qquad
z =(z^0, \vec{z}) \in {\cal V}_{-}, \end{equation}
where  ${\cal V}_{-}
\equiv \{(z^0, \vec{z}) \in R^n | z^0 < - |\vec{z}| \}$ is the lower light
cone. The limit (1.5) implies  $z^0 \rightarrow  -\infty$ and
under certain additional assumptions describes
the approaching to the singularity.

\section{Billiard  representation}
\setcounter{equation}{0}

We describe briefly our recent results on billiard representation
near the "singularity" for the dynamical system (1.1)  [10,11].
We restrict the Lagrange system (1.1) on the lower
light cone  and introduce the analogues of
the Misner-Chitre coordinates in
$\cal{V}_{-}$ [12,13]
\begin{eqnarray} 
&&z^0 = - \exp(-y^0) \frac{1 + \vec{y}^2}{1 - \vec{y}^2}, \\
&&\vec{z} = - 2 \exp(-y^0) \frac{ \vec{y}}{1 - \vec{y}^2},
\end{eqnarray}
$|\vec{y}| < 1$. We also fix the gauge (or time parametrization)
\be 
{\cal N} = \exp(- 2 y^0) = - z^2.
\ee
We get for the restriction of the Lagrangian  on ${\cal V}_{-}$
\begin{equation} 
L_{-} = \frac{1}{2}
[- (\dot{y}^{0})^2 + {h_{ij}}(\vec{y}) \dot{y}^{i} \dot{y}^{j}]
-  \exp(- 2y^0) V.
\end{equation}
Here
\be 
{h_{ij}}(\vec{y}) = 4 \delta_{ij} (1 - \vec{y}^2)^{-2},
\ee
$i,j =1, \ldots , n-1$, are the components of the Riemannian
metric on the $(n-1)$-dimensional open unit disk (ball)
\be 
D^{n-1} \equiv \{ \vec{y}= (y^1, \ldots, y^{n-1})| |\vec{y}| < 1 \}
\subset R^{n-1}. \ee
The pair $(D^{n-1}, h = {h_{ij}}(\vec{y}) dy^i \otimes dy^j)$ is one
of the realizations of the $(n-1)$-dimensional Lobachevsky space
$H^{n-1}$. We also get the energy constraint
\be 
E_{-} = - \frac{1}{2}  (\dot{y}^{0})^2 +  \frac{1}{2}
{h_{ij}}(\vec{y}) \dot{y}^{i} \dot{y}^{j} +  \exp(- 2y^0) V = 0.
\ee

Now we are interested in the behavior of the dynamical system
in the limit  $y^0 \rightarrow - \infty$ (or, equivalently, in
the limit (1.5)).
Using the restrictions (1.3), (1.4)
we obtain [11]
\ba 
\lim_{y^0 \rightarrow - \infty} \exp(- 2y^0) (V - A_0) =
{V}(\vec{y},B) \equiv &0, &\vec{y} \in B, \nonumber \\
&+ \infty, &\vec{y} \in D^{n-1} \setminus B,
\ea
where
\be 
B = \bigcap_{\alpha \in
\Delta_{+}} {B}(u^{\alpha})  \subset D^{n-1}, \ee
\be 
\Delta_{+}  \equiv \{ \alpha | (u^{\alpha})^2 > 0 \},  \ee
and
\be 
{B}(u^{\alpha}) \equiv  \{ \vec{y} \in D^{n-1} |
|\vec{y} + \frac{\vec{u}^{\alpha}}{u_{0}^{\alpha}}| >
\sqrt{(\frac{\vec{u}^{\alpha}}{u_0^{\alpha}})^2 - 1} \},
\ee
$\alpha \in \Delta_{+}$. $B$ is an open domain.
Its boundary $\partial B = \bar{B} \setminus B$ is formed by
certain parts of $m_{+} = |\Delta_{+}|$
($m_{+}$ is the number of elements in $\Delta_{+}$) of $(n-2)$-dimensional
spheres with the centers in the points
\be 
\vec{v}^{\alpha} = - \vec{u}^{\alpha}/u^{\alpha}_{0}, \qquad
\alpha \in \Delta_{+},
\ee
($|\vec{v^{\alpha}}| > 1$) and radii
\be 
r_{\alpha} = \sqrt{(\vec{v}^{\alpha})^2 - 1}
\ee
respectively.
So, in the limit $y^{0} \rightarrow - \infty$ we are led to the
dynamical system
\ba 
&L_{\infty} = - \frac{1}{2} (\dot{y}^{0})^2 +  \frac{1}{2}
{h_{ij}}(\vec{y}) \dot{y}^{i} \dot{y}^{j} -  V_{\infty}, \\
&E_{\infty} = - \frac{1}{2} (\dot{y}^{0})^2 +  \frac{1}{2}
{h_{ij}}(\vec{y}) \dot{y}^{i} \dot{y}^{j} +  V_{\infty} = 0,
\ea
where
\be  
V_{\infty} = A_0  \exp(- 2y^0) + {V}(\vec{y},B).
\ee
After the separating of $y^0$ variable
\ba
y^0 = && \omega (t - t_0), \qquad  A_0 = 0,   \nonumber   \\
      &&  \frac{1}{2} \ln[\frac{2 A_0}{\omega^2} \sinh^2(\omega (t -
               t_0)), \, A_0 \neq 0. \ea
($\omega > 0$ , $t_0$  are constants) the dynamical system
is reduced to the
Lagrange system with the Lagrangian
\be 
L_{B} =  \frac{1}{2}
{h_{ij}}(\vec{y}) \dot{y}^{i} \dot{y}^{j} -  {V}(\vec{y},B). \ee
Due to (2.17)
\be 
E_{B} =  \frac{1}{2} {h_{ij}}(\vec{y}) \dot{y}^{i}
\dot{y}^{j} +  {V}(\vec{y},B) = \frac{\omega^2}{2}.
\ee
The limits
$t \rightarrow - \infty$   for $A_0 = 0$ and
$t \rightarrow t_0 + 0$   for $A_0 \neq 0$ describe the approach to the
singularity.  When the set (2.10) is empty ($\Delta_{+} = \emptyset$)  we
have $B = D^{n-1}$ and the Lagrangian (2.18)  describes  the  geodesic
flow  on  the Lobachevsky space $H^{n-1} = (D^{n-1}, h_{ij} dy^i \otimes
dy^j)$.

When  $\Delta_{+} \neq \emptyset$ the Lagrangian
(2.18) describes the motion of the particle  of  unit  mass,  moving
in the ($n-1$)-dimensional billiard $B \subset D^{n-1}$  (see (2.9)).
The geodesic motion in $B$  corresponds to a "Kasner epoch" and the
reflection from the boundary corresponds to the change of Kasner epochs.
Let $A_0 = 0$. When the volume of $B$ is finite:
$volB < + \infty$, we have a stochastic behaviour near the singularity.
Such situation takes place in Bianchi-IX cosmology [2].
When the billiard $B$  has an infinite volume: $vol B = +\infty$
there are  open zones of non-zero measure at  the  infinite  sphere
$S^{n-2} = \{|\vec{y}| =1 \}$. After a finite number of reflections from
the boundary the  particle  moves  toward  one  of  these  open  zones.
For corresponding cosmological model we get the "Kasner-like" behavior in
the limit $t \rightarrow - \infty$.
For $A_0 \neq 0$ (when Zeldovich matter is present)
in the limit $t \rightarrow t_0 + 0$
we have $y^0 \rightarrow - \infty$ and  ${\vec{y}}(t)
\rightarrow \vec{y}_0 \in B$. So, the stochastic behavior near the
singularity is absent in this case.

Proposition [11]. The billiard $B$ (2.9) has a finite volume if and only if
the point-like sources of light located at the points $\vec{v}^{\alpha}$
(2.12) illuminate the unit sphere $S^{n-2}$.

The problem of illumination of convex body in
multidimensional vector space by point-like sources for the first time was
considered in [14, 15]. For the case of $S^{n-2}$ this problem is
equivalent to the problem of covering the spheres with spheres.  There
exist a topological bound on the number of point-like sources $m_{+}$
illuminating the sphere $S^{n-2}$ [15]:
\be 
m_{+} \geq n.
\ee
So, the stochastic behaviour for the solutions of Lagrange equations
for the Lagrangian (1.1) with the gauge fixing (2.3)
the limit (1.5)
(near the "singularity") may take place only if $A_0 = 0$ and
the number of terms in the potential with $(u^{\alpha})^2 > 0 $ is
no less than the minisuperspace dimension.

\section{Quantum case}
\setcounter{equation}{0}

The quantization of zero-energy constraint (2.7) leads to the
Wheeler-DeWitt (WDW) equation in the gauge (2.3) [16,17]
\begin{equation} 
(- \frac{1}{2} {\Delta}[\bar{G}]  + a_{n}
{R}[\bar{G}]  +  \exp(- 2y^0) V ) \Psi =0.
\end{equation}
Here
$\Psi={\Psi}(y)$ is "the wave function of the Universe",
$V = {V}(y)$ is the
potential (1.2), $a_{n} = (n-2)/8(n-1)$,
${\Delta}[\bar{G}]$ and ${R}[\bar{G}]$ are the Laplace-Beltrami
operator and the scalar curvature of the minisuperspace metric
\begin{equation} 
\bar{G} = - dy^0 \otimes dy^0 + h, \qquad
 h = {h_{ij}}(\vec{y}) dy^i \otimes dy^j.
\end{equation}
(We remind that, $h$ is the metric on Lobachevsky space $D^{n-1}$.)
The form of WDW eq. (3.1) follows from the demands of minisuperspace
invariance and conformal covariance. Using relations
\begin{equation} 
{\Delta}[\bar{G}]  = - (\frac{\partial}{\partial y^0})^2 + {\Delta}[h],
\qquad {R}[\bar{G}] =   {R}[h] = - (n-1)(n-2),
\ee
we rewrite (3.1) in the form
\begin{equation} 
(\frac{1}{2} (\frac{\partial}{\partial y^0})^2 - \frac{1}{2} {\Delta}[h] -
\frac{(n-2)^2}{8}  + \exp(- 2y^0) V) \Psi =0.
\end{equation} In the limit $y^0
\rightarrow - \infty$ the WDW eq. reduces to the relations
\begin{equation} 
((\frac{\partial}{\partial y^0})^2 + 2 A_0 \exp(- 2y^0)
- {\Delta_{*}}[h]) \Psi_{\infty} = 0,
\qquad \Psi_{\infty}|_{\partial B} = 0,
\ee where $\partial B = \bar B
\setminus B$ is the boundary of the billiard $B$ (2.9) (in $D^{n-1}$) and
\begin{equation} 
{\Delta_{*}}[h] =  {\Delta}[h] + \frac{(n-2)^2}{4}.
\ee
Now, we suppose that $\bar{B}$ is compact and the operator (3.6)
with the boundary condition (3.5) has a negative spectrum, i.e.
\begin{equation} 
{\Delta_{*}}[h] u_{n} = - \omega_{n}^2  u_{n},
\ee
$\omega_n > 0$, $n = 0,1, \ldots$ (this is valid at least for "small
enough" $B$). Using (3.7) we get the general solution
of the asymptotic WDW eq. (3.5)
\begin{equation} 
{\Psi_{\infty}}(y^0, \vec{y}) = \sum_{n=0}^{\infty}
[a_n {\Psi_{n}}(y^0, \vec{y}) +
a_n^{*} {\Psi_{n}^{*}}(y^0, \vec{y})].
\ee
Here
\be 
{\Psi_{n}}(y^0, \vec{y}) = {f_{n}}(y^0) {u_{n}}(\vec{y}),
\ee
where the set ${u_n}$ in (3.7) is orthonormal basis in $L_2(B)$:
\be 
(u_n, u_m) = \delta_{mn},
\ee
$m, n = 0,1, \ldots$. For $A_0 = 0$
\be 
{f_{n}}(t) = (2 \omega_n)^{-1/2} \exp( -i \omega_n y^0).
\ee
The second quantized model (in gravity also callled as "third
quantized") in this case is defined by
eq. (3.8) with the familiar relations imposed
\be 
[a_n, a_m^*] = \delta_{mn}, \qquad a_m^* = (a_m)^+ , \qquad a_n |0> = 0,
\ee
$m, n = 0,1, \ldots$. For $A_0 \neq 0$ we consider two basis
\ba  
&&{f_{n}^{in}}(y^0) = [\frac{\pi}{2 \sinh(\pi \omega_n)}]^{1/2}
{J_{-i\omega_n}}(\sqrt{2A_0} e^{-y^0}), \\
&&{f_{n}^{out}}(y^0) = \frac{1}{2} (\pi)^{1/2}  \exp(- \pi \omega_n /2)
{H^{(1)}_{i\omega_n}}(\sqrt{2A_0} e^{-y^0}),
\ea
where $J_{\nu}$, $H^{(1)}_{\nu}$  are Bessel and Hankel
functions respectively. The functions (3.13), (3.14) are related by
Bogoljubov transformation [18]
\be
f_{n}^{out} = \alpha_n  f_{n}^{in} + \beta_n (f_{n}^{in})^{*},
\ee
where
\be
\alpha_n = (\frac{e^{\pi \omega_n}}{2 \sinh \pi \omega_n})^{1/2}, \qquad
\beta_n = - (\frac{e^{- \pi \omega_n}}{2 \sinh \pi \omega_n})^{1/2}.
\ee
A standard calculation [18] gives the mean number of "out-Universes"
containing in "in-vacuum"
\be
N_n = <0,in| a^{+}_{n,out} a_{n,out} |0, in> =
|\beta_n|^{2} = (\exp(2 \pi \omega_n) - 1)^{-1}.
\ee
So, we obtained the Planck distribution with the temperature
$T = 1/ 2 \pi$. We note that for Bianchi-IX case the considered scheme was
suggested in [19].

\pagebreak

\pagebreak

\end{document}